\documentclass[useAMS,usenatbib]{mnras}

\usepackage{graphicx}
\usepackage{txfonts}

\title[Propagation of star formation]
      {Propagation of star formation at sub-kiloparsec scales}

\author[A.~S.~Gusev \& E.~V.~Shimanovskaya]
       {A.~S.~Gusev 
        and E.~V.~Shimanovskaya \\
        Sternberg Astronomical Institute, Lomonosov Moscow State University, 
        Universitetsky pr. 13, 119992 Moscow, Russia \\
             }

\date{Accepted 2019 June 29. Received 2019 June 29; 
in original form 2019 March 5}

\begin{document}

\maketitle

\begin{abstract}
We study the propagation of star formation based on the investigation of 
the separation of young star clusters from H\,{\sc ii} regions nearest 
to them. The relation between the separation and $U-B$ colour index 
(or age) of a star cluster was found. The average age of star clusters 
increases with the separation as the 1.0-1.2 power in the separation 
range from 40 to 200 pc and as the 0.4-0.9 power in the range of 100-500 pc 
in the galaxies with symmetric morphology. The galaxies with distorted 
asymmetric disc structure show more complex and steeper (power $>1.2$ at 
separations from 40 to 500 pc) dependence between the age and the 
separation. Our results confirm the findings of previous studies on the 
dominant role of turbulence in propagation of the star formation process 
on spatial scales up to 500 pc and on time scales up to 300 Myr. On a smaller 
scale ($\le100$ pc), other physical processes, such as stellar winds and 
supernova explosions, play an important role along with turbulence. On 
the scale of stellar associations (100-200 pc and smaller), the velocity 
of star formation propagation is almost constant and it has a typical 
value of a few km\,s$^{-1}$.
\end{abstract}


\section{Introduction}

Physical processes such as gravitational collapse and turbulence 
compression are believed to play a key role in the evolution of star 
formation regions over a wide range of scales, from smallest clumps of 
young stars to star complexes \citep*{scalo1985,elmegreen2010}. 
The sequential hierarchical structures of young stellar groups in 
these regions are explained in terms of triggered star formation, when 
new stars form as the result of gravitational instabilities in  
shocked layers initiated by expanding H\,{\sc ii} bubbles 
\citep*{elmegreen1977,efremov1998,elmegreen2000,elmegreen2002,
elmegreen2006,elmegreen2011}. The largest coherent star formation 
regions are star complexes with diameters of the order of $\sim500-600$ pc 
\citep*{elmegreen1996,efremov1998}.

Massive star formation complexes in turbulent discs are the result 
of instabilities in self-gravitating gas discs \citep{fisher2017}. They 
are observed in both nearby and distant galaxies. Large star formation 
regions in galaxies at $z\sim1-2$ are seen as star formation clumps in 
H$\alpha$ \citep{ali2017,fisher2017}. Sizes of these complexes/clumps 
correlate with turbulence and gas fraction in discs \citep{fisher2017}. 
Modern methods of analysis of the largest coherent star formation scales 
such as two-point correlation function give results which are in 
good agreement with both classic methods (Gaussians, for instance) and 
theoretical Jeans lengths \citep{ali2017}.

The turbulence predicts a relation $t\sim S^{0.5}$ between age $t$ and 
size $S$ of star formation regions \citep{elmegreen1996,efremov1998}. 
A diffusion-driven expansion would produce a relation $t\sim S^2$ 
between age and size \citep{grasha2017}. \citet{elmegreen1996} 
and \citet{efremov1998} originally studied positions and ages of Cepheid 
variables and star clusters in the Large Magellanic Cloud and found that 
the average age difference between pairs of star clusters increases with 
their separation as $\Delta t {\rm (Myr)}\sim 3.3S {\rm (pc)}^{0.33}$ in 
the separation range from 8 to 780 pc. Later \citet{marcos2009} found 
that the Milky Way star cluster pairs also satisfy the relation between 
age differences and their separation with the $0.40\pm0.08$ power if the 
effects of incompleteness and cluster dissolution are taken into account. 
\citet{grasha2017} studied the connection between the age difference and 
separation of star cluster pairs in eight local galaxies and found 
that the age difference increases with the cluster pair separation to 
the $0.25-0.6$ power, with the maximum size, over which star formation 
is physically correlated, ranging from $\sim200$ to $\sim1000$ pc in 
different galaxies. Numerical simulations investigating the 
propagation of star formation in a turbulent medium derive an 
age-separation relation to the 0.5 power for star groups with 
separations $>50$ pc \citep{nomura2001}, which is in agreement with the 
observational data.

The relation between age difference and separation for star clusters is 
also similar to the correlation between size and line width of giant 
molecular clouds (GMC) suggesting that the crossing time in a GMC 
increases as the square root of the size of the star formation region 
\citep{larson1981,elmegreen1996}.

In this article, we develop an alternative method for studying processes 
of triggered star formation. In \citet{gusev2018} we found, but only 
noted a dependence between separations of star clusters from the nearest 
H\,{\sc ii} region and $U-B$ colour index of the star cluster for the 
pairs with non-deprojected angular separation from 1.5 to 7 arcsec. This 
study is devoted to the analysis of this dependence. As is known, $U-B$ 
colour index can be used as an age indicator for young stellar systems 
\citep{whitmore2010}.

The youngest star formation regions with ages less than 10 Myr are usually 
associated with H\,{\sc ii} regions. They evolves during the first 
tens Myr of their lives through several evolutionary stages, from the stage 
when young stars clumps are completely obscured by their dusty gas 
cocoons to the stage of a young star cluster with no evidence of 
H\,{\sc ii} \citep{lada2003}. \citet{whitmore2011} developed 
the evolutionary classification scheme of star clusters based on the
{\it Hubble Space Telescope (HST)} observations of M83. Star clusters 
become visible in optical bands since about 1~Myr ('obscured clusters'). 
Ionized gas stays spatially coincident with cluster stars during 
$1-4$ Myr ('emerging clusters'). Clusters with ages $\approx4-5$ Myr 
have a small H\,{\sc ii} bubble with radius of $7-20$ pc ('very young 
clusters'), surrounding the cluster. Star clusters with ages $\ge 5$ Myr 
have a larger ionized gas bubble surrounding the cluster 
('young clusters'). H\,{\sc ii} bubbles are dissolved around star
formation regions older than $8-10$ Myr. From the age $>10$ Myr, we observe 
blue, but redden ageing star clusters without H$\alpha$ emission
\citep['intermediate-age clusters' by][]{whitmore2011}. These clusters are 
free from dust.

Thus, the dependence between separations of star clusters from the nearest 
H$\alpha$ source and $U-B$ colour index (age) of a star cluster 
can be used to directly measure the velocity of star formation 
propagation.

\section{Search and selection of star formation regions}

\subsection{Observations and data reduction}

The sample of galaxies is based on our photometric survey of 26 galaxies 
\citep{gusev2015}. We selected nine of them (NGC 266, NGC 628, NGC 2336, 
NGC 3184, NGC 3726, NGC 5585, NGC 6217, NGC 6946, and NGC 7331) with 
observed star formation regions and available H$\alpha$ 
spectrophotometric data.

Our own photometric $UBVRI$ and H$\alpha$ (H$\alpha$+[N\,{\sc ii}]) 
observations were described and published earlier 
\citep[see][and references therein]{gusev2018}.

\begin{figure}
\vspace{0.3cm}
\resizebox{0.95\hsize}{!}{\includegraphics[angle=000]{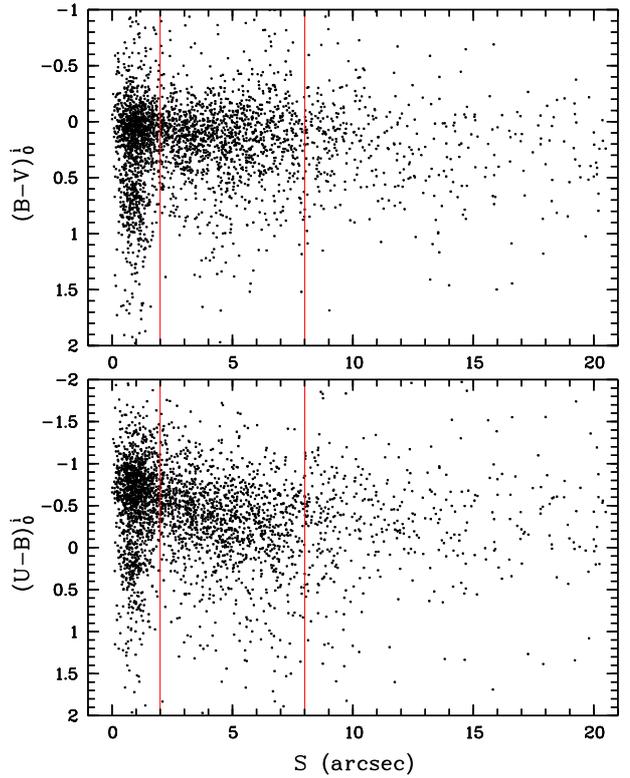}}
\caption{Colour indices $B-V$ (top) and $U-B$ (bottom) of star clusters, 
corrected for Galactic extinction and inclination effects, vs. its 
deprojected separation from the nearest H\,{\sc ii} region. Vertical 
lines confine the area between 2 and 8 arcsec, where the correlation 
between $U-B$ and a separation is observed.
}
\label{figure:ubv_sep}
\end{figure}

We did not observe NGC~3184 from our sample in $U$ filter. Therefore we 
used an $U$ image of NGC~3184 obtained by \citet{larsen1999b} and 
available in the NED\footnote{\url{http://ned.ipac.caltech.edu/}} 
database. The absolute calibration of the $U$ image of the galaxy was 
carried out using parameters of descriptors from original 
FITS files (images) and their explanations in \citet{larsen1999b}. 
Additionally, we checked the calibration using the aperture photometry 
data for the galaxies from the 
LEDA\footnote{\url{http://leda.univ-lyon1.fr/}} database 
\citep{paturel2003}. We also compared our $BVRI$ images of NGC~3184 
with those from \citet{larsen1999b}. Pixel-to-pixel deviations are 
within the stated accuracy of measurements and they did not show any 
systematic errors.

We also used H$\alpha$+[N\,{\sc ii}] FITS images of NGC~3726 and NGC~5585 
obtained by \citet{knapen2004} and \citet{dale2009}. Parameters for 
absolute calibration of H$\alpha$+[N\,{\sc ii}] flux to units of 
erg\,s$^{-1}$cm$^{-2}$ were found in descriptors of the FITS files. 
Additionally, we checked the calibration using the integrated 
H$\alpha$+[N\,{\sc ii}] fluxes of NGC~5585 measured in \citet{james2004} 
and \citet{kennicutt2008}.

These data were used for identification and selection of star 
formation regions. 

\begin{table*}
\caption[]{\label{table:sample}
The galaxy sample.
}
\begin{center}
\begin{tabular}{ccccccccccc} \hline \hline
Galaxy & Type & $B_t$ & $M_B^a$ & Inclination & PA       & $R_{25}^b$ & 
$R_{25}^b$ & $D$   & $A(B)_{\rm Gal}$ & $A(B)_{\rm in}$ \\
       &      & (mag) & (mag)   & (degree)    & (degree) & (arcmin)   & 
(kpc)      & (Mpc) & (mag)        & (mag) \\
1 & 2 & 3 & 4 & 5 & 6 & 7 & 8 & 9 & 10 & 11 \\
\hline
NGC~628   & SA(s)c      &  9.70 & --20.72 &  7 &  25 & 5.23 & 11.0 &  7.2
& 0.254 & 0.04 \\
NGC~3184  & SAB(rs)cd   & 10.31 & --19.98 & 14 & 117 & 3.79 & 11.3 & 10.2
& 0.060 & 0.02 \\
NGC~3726  & SAB(r)c     & 10.31 & --20.72 & 49 &  16 & 2.62 & 10.9 & 14.3
& 0.060 & 0.30 \\
NGC~5585  & SAB(s)d     & 10.94 & --18.73 & 53 &  34 & 2.13 &  3.5 &  5.7
& 0.057 & 0.38 \\
NGC~6946  & SAB(rs)cd   &  9.75 & --20.68 & 31 &  62 & 7.74 & 13.3 &  5.9
& 1.241 & 0.04 \\
\hline
\end{tabular}
\end{center}
\begin{flushleft}
$^a$ Absolute magnitude of a galaxy corrected for Galactic extinction and 
inclination effects. \\
$^b$ Radius of a galaxy at the isophotal level 25 mag/arcsec$^2$ in the 
$B$ band corrected for Galactic extinction and inclination effects. \\
\end{flushleft}
\end{table*}

\subsection{Selection of young star clusters and H\,{\sc ii} regions}
\label{sect:select}

Procedure of identification and preliminary selection of star 
formation regions on $B$ and H$\alpha$ images of galaxies using the 
{\sc SExtractor}\footnote{\url{http://sextractor.sourceforge.net/}} 
program was described in detail in \citet{gusev2018}.

Coordinates of all objects (star clusters on $B$ images and 
H\,{\sc ii} regions on H$\alpha$ images) were recalculated 
for the deprojected positions using position angles and inclinations 
of the galaxies by standard formulae. In the next step, we determined 
the separation, $S$, between a star cluster and the nearest 
H\,{\sc ii} region (SC-H{\sc ii}R pair) for all selected star clusters. 
Herewith, we suggest that a single H\,{\sc ii} region may be 
paired with up to multiple star clusters.

The relation between colour indices of young star clusters and their 
separations is given in Fig.~\ref{figure:ubv_sep}. Unlike 
Fig.~3 in \citet{gusev2018}, Fig.~\ref{figure:ubv_sep} shows 
deprojected angular separations.

\citet{whitmore2011} showed that the morphology of ionized gas 
bubbles can be used as an age indicator of star clusters. Unfortunately, 
we are not able to study the H\,{\sc ii} morphology directly on scales of 
$<40$ pc, because our angular resolution is 1-1.5 arcsec. However, as was 
shown in \citet{gusev2016}, the distance between photometric centres of 
gas emission and star clusters correlates within a single star formation 
region with the relation between stellar and gas absorption, depending on 
the age, and can be used as an evolutionary indicator.

The diagrams in Fig.~\ref{figure:ubv_sep} show that clusters with close 
H$\alpha$ emission sources ($S<$ 2 arcsec) have a wide range of colours. 
This is a result of a significant absorption in young dusty star clusters.

Fig.~\ref{figure:ubv_sep} shows a correlation between the separating 
distance from the nearest H\,{\sc ii} region and $U-B$ colour index of a
star cluster: clusters with 2 arcsec $<S<$ 8 arcsec become redder 
with increase of separations. Here we do not observe objects with  
extremely red colour indices. Apparently, the absorption in these star 
clusters without nearby H$\alpha$ emission is close to zero 
\citep{whitmore2011}.

\begin{figure}
\vspace{0.3cm}
\resizebox{0.95\hsize}{!}{\includegraphics[angle=000]{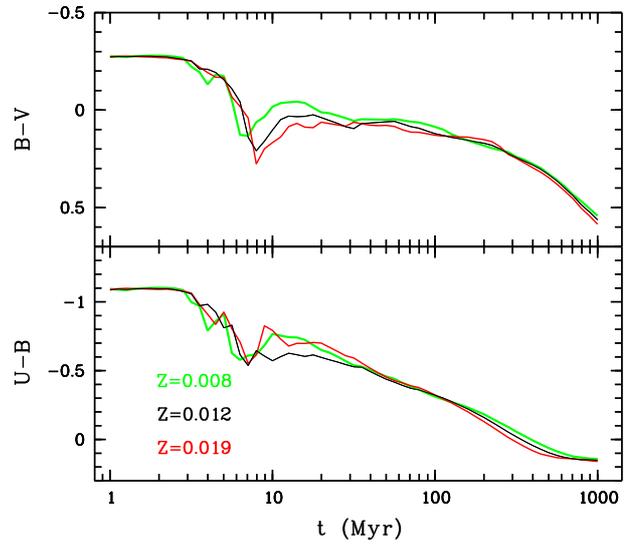}}
\caption{Variations of colour indices $B-V$ (top) and $U-B$ (bottom) 
of synthetic stellar systems with $Z=0.008$, 0.012, and 0.019 vs. age. 
}
\label{figure:ubv_age}
\end{figure}

Remark the absence of any dependence between separation and 
$B-V$ colour index for star clusters with 2 arcsec $<S<$ 8 arcsec 
(Fig.~\ref{figure:ubv_sep}). The reason is that young stellar 
systems with ages from 10 to 200-300 Myr have an approximately constant 
$B-V$ colour index unlike $U-B$. It is clearly illustrated in 
Fig.~\ref{figure:ubv_age} where simple stellar population (SSP) 
models with Salpeter IMF and the interval of masses, ranging from
$0.1 M_{\odot}$ to $100 M_{\odot}$, are shown. Here we used the grid 
of isochrones provided by the Padova group \citep{marigo2008} 
(version 2.3) and obtained via the online server 
CMD\footnote{\url{http://stev.oapd.inaf.it/cgi-bin/cmd}}.

We did not find a visual correlation between separation from 
H\,{\sc ii}~regions and colours of star clusters for objects with 
SC-H{\sc ii}R deprojected separation larger than 8~arcsec.

\citet{efremov1998} estimated a duration of star formation 
in star complexes at large spatial scales (500-1000~pc) to be about 30~Myr. 
Among star clusters within a complex, both very young ($<10$~Myr) 
and intermediate-age objects ($\sim100$~Myr) are encountered. The colour 
indices of star clusters in Fig.~\ref{figure:ubv_sep} correspond to 
the range from several Myr to a few hundred Myr (see 
Fig.~\ref{figure:ubv_age}). 

Our angular resolution does not allow us to study stellar groups at 
the scale of star clusters in distant galaxies. For larger star formation 
regions, SSP model is not correct \citep{efremov1998} and the colours of 
these stellar groups depend on both the age and the star formation 
history. For this reason we eliminated three distant ($D>$ 20 Mpc) galaxies 
NGC 266, NGC 2336, and NGC 6217 from further consideration. We also 
excluded the high inclination galaxy NGC 7331 ($i=75\degr$), because 
of large uncertainty of projection effects.

\begin{figure}
\vspace{0.6cm}
\resizebox{0.90\hsize}{!}{\includegraphics[angle=000]{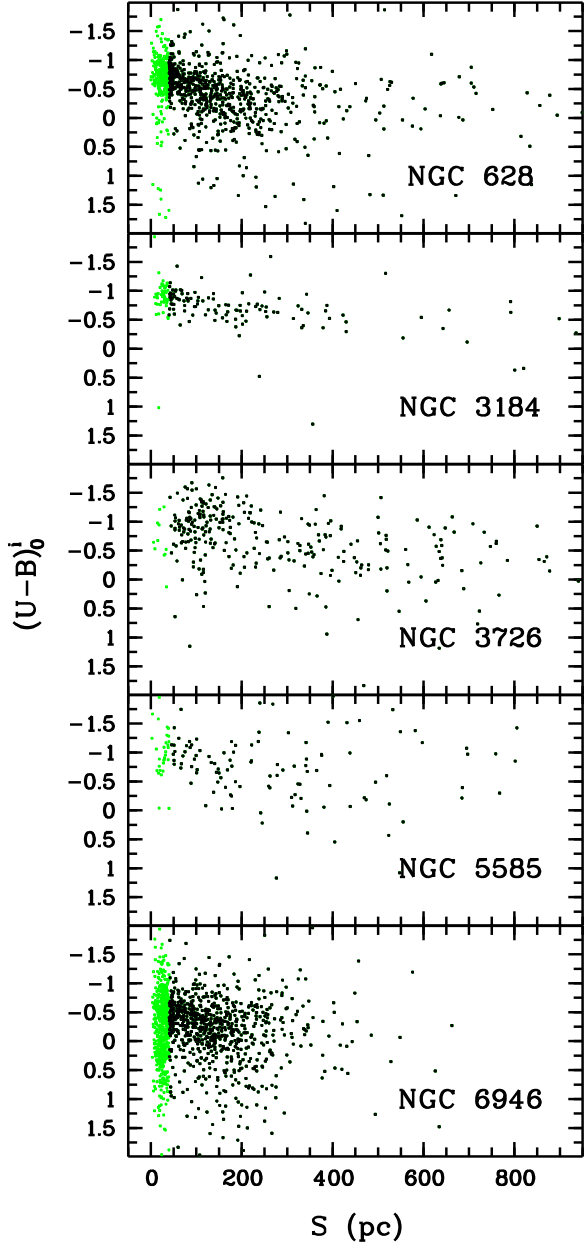}}
\caption{The colour index $U-B$ of star clusters, corrected for Galactic 
extinction and inclination effects, vs. a linear separation of 
SC-H{\sc ii}R pairs. Green dots correspond to star clusters with a linear 
separation $S<40$~pc.
}
\label{figure:col_dist}
\end{figure}

The sample of remaining five galaxies is presented in 
Table~\ref{table:sample}, where the Galactic absorption, 
$A(B)_{\rm Gal}$, is taken from the NED database, and  
other parameters are taken from the LEDA database. 
The morphological type of a galaxy is shown 
in column (2), the apparent and absolute magnitudes are given in columns 
(3) and (4), the inclination and position angles are listed in columns (5) 
and (6), and the isophotal radii in arcmin and kpc are presented in 
columns (7) and (8). The adopted distances are given in column (9). The 
Galactic absorption and the dust absorption due to the inclination of a 
galaxy are presented in columns (10) and (11). The adopted value of the 
Hubble constant is equal to H$_0 = 75$ km\,s$^{-1}$Mpc$^{-1}$.

Fig.~\ref{figure:col_dist} shows dependences between the $U-B$ colour 
indexof star clusters and the linear separation of SC-H{\sc ii}R pairs 
for five studied galaxies. Here we excluded star clusters in close 
SC-H{\sc ii}R pairs with undefined absorption ($S<40$ pc, green dots 
in Fig.~\ref{figure:col_dist}) and clusters with large measured 
photometric errors ($\Delta U>0.3$ mag and/or $\Delta B>0.3$ mag). 
The remaining star clusters have no H$\alpha$ emission and therefore 
they are older than 10 Myr. As seen from the figure, a correlation 
between $U-B$ and separation is observed on distances up to 300 pc and 
more. Below we will consider the star clusters in the range of separation 
where the linear correlation between colour and separation is observed.

After preliminary selection, the sample included 741 pair 
candidates in NGC 628, 116 pairs in NGC 3184, 297 pairs in NGC 3726, 
126 pairs in NGC 5585, and 808 pairs in NGC 6946.

\begin{figure}
\vspace{0.6cm}
\resizebox{0.85\hsize}{!}{\includegraphics[angle=000]{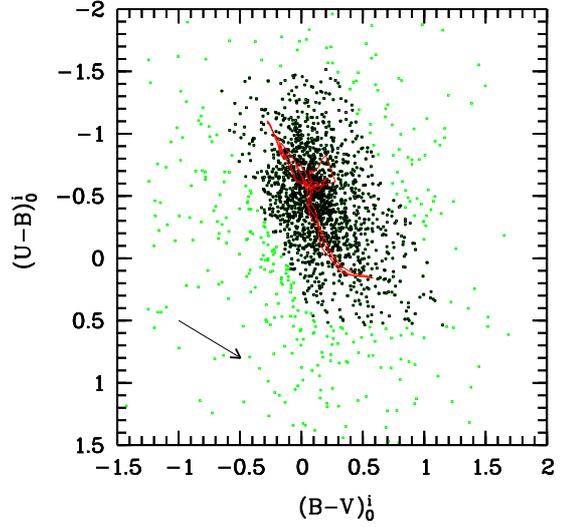}}
\caption{Colour-colour diagram $(U-B)-(B-V)$ for star clusters. Red 
curves represent SSP CMD models with continuously populated IMF in 
the age interval from 1~Myr to 1~Gyr with Salpeter's slope 
$\alpha = -2.35$ and $Z = 0.008$, 0.012, and 0.019. Green dots correspond 
to objects with colour indices outside the area of star clusters models. 
The reddening vector is shown. See the text for details.
}
\label{figure:ub_logs}
\end{figure}

\begin{figure}
\vspace{0.6cm}
\resizebox{0.85\hsize}{!}{\includegraphics[angle=000]{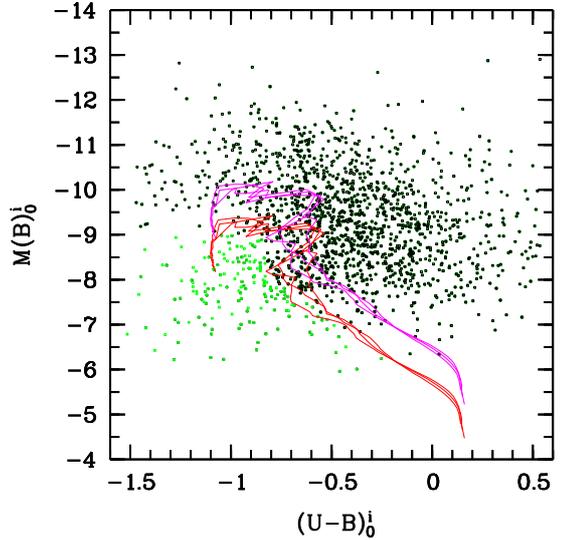}}
\caption{The colour-magnitude diagram for star clusters. Evolutionary 
tracks for models with Salpeter's IMF and $Z=0.008$, 0.012, and 
0.019, drawn in the age interval from 1~Myr to 1~Gyr, are shown. The 
tracks were computed for star clusters with masses $10^4 M_\odot$ 
(magenta curves) and $5\cdot10^3 M_\odot$ (red curves). Green dots 
correspond to star clusters with masses less than 
$5\cdot10^3 M_\odot$. See the text for details.
}
\label{figure:ub_logs2}
\end{figure}

A native source extractor selection is known to include a significant 
number of non-cluster contaminants after generating the star clusters data 
base \citep{grasha2015}. A good test for estimation of the contamination of 
our star cluster catalogue and its clearing from non-cluster objects, 
such as background stars and galaxies, and cosmic rays, are $U-B$ colours 
\citep{grasha2019}. We checked the disposition of our star cluster 
candidates on the $(U-B)-(B-V)$ colour-colour diagram 
(Fig.~\ref{figure:ub_logs}).

We left for further consideration star cluster candidates whose colours, 
within maximum allowable errors, are consistent with the colours of SSP 
CMD evolutionary tracks for stellar systems younger than 1 Gyr (black dots 
in Fig.~\ref{figure:ub_logs}). The red limit in $(U-B)^i_0=0.54$ mag is also 
the boundary which separates young star clusters with $B-V\sim0.5$ mag from 
the background Main Sequence yellow stars \citep{whitmore2010}.

A direct unambiguous relation between age and $U-B$ colour index of a stellar 
system is possible only in terms of continuously populated IMF. Stochastic 
effects in the discrete randomly populated IMF begin to play a key role for 
SSP stellar systems (star clusters) with masses less than 
$5\cdot10^3-10^4 M_\odot$ \citep{whitmore2010,piskunov2011}. This effect 
is clearly illustrated in Fig.~19 in \citet{gusev2016}. Due to the 
impossibility of finding an age by $U-B$ colour for low-massive 
clusters, we excluded star clusters with masses $<5\cdot 10^3 M_\odot$ 
from further consideration (green dots in Fig.~\ref{figure:ub_logs2}).

\begin{figure}
\resizebox{1.00\hsize}{!}{\includegraphics[angle=000]{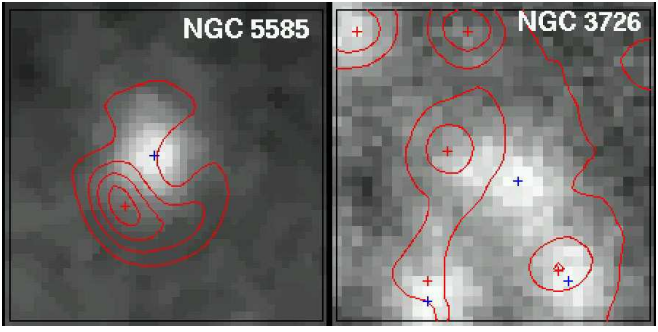}}
\caption{$B$ images of selected star clusters in NGC 5585 and NGC 3726 
with overlaid isophotes in the H$\alpha$ line. The angular size of 
both images is $11.7 \times 11.7$ arcsec$^2$, which corresponds to the 
linear size of 320 pc in NGC 5585 and 810 pc in NGC 3726. The centres of 
star clusters (blue crosses) and H\,{\sc ii}~regions (red crosses) are 
shown. North is upwards and east is to the left.
}
\label{figure:map1}
\end{figure}

\begin{figure}
\resizebox{1.00\hsize}{!}{\includegraphics[angle=000]{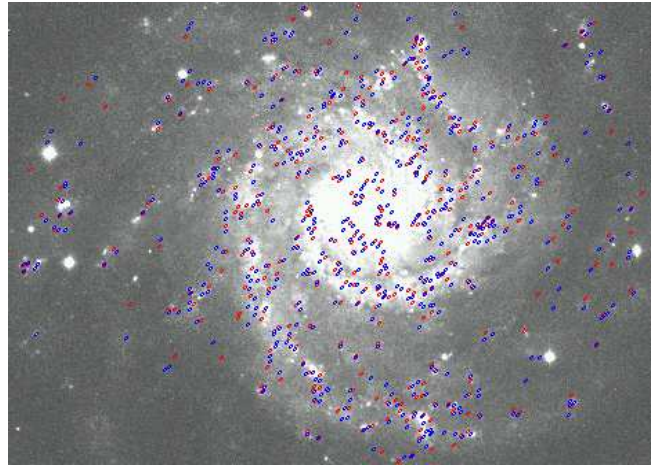}}
\caption{$B$ image of NGC 628 and positions of selected star 
clusters (blue dots) and H\,{\sc ii}~regions (red dots) in the galaxy. 
North is upwards and east is to the left.
}
\label{figure:map2}
\end{figure}

The final sample includes 503 SC-H{\sc ii}R pairs in NGC 628, 70 
pairs in NGC 3184, 254 pairs in NGC 3726, 40 pairs in NGC 5585, and 577 
pairs in NGC 6946. To illustrate the selected objects, we present a couple 
of postage stamp images of H\,{\sc ii} regions and star clusters in the 
closest (NGC 5585) and the most distant (NGC 3726) galaxies in 
Fig.~\ref{figure:map1}. Our full sample of identified and finally selected 
star clusters and H\,{\sc ii} regions is shown in Fig.~\ref{figure:map2} 
for the NGC 628 galaxy.

\section{Results}

\subsection{Cluster age versus separation relation}
\label{sect:age_sep}

Fig.~\ref{figure:col_dist} shows a clear linear dependence between the 
colour index $U-B$ and the separation within a wide range 
of $S$ for all five studied galaxies.

Note here that typical $U-B$ colours differ slightly from 
galaxy to galaxy (Fig.~\ref{figure:col_dist}). This may be a result 
of zero-point calibration errors and/or uncertainty of Galactic or 
internal extinctions. A special case is NGC 6946 which is located on 
the low Galactic latitude (see column (10) in Table~\ref{table:sample}).

Unlike \citet{efremov1998} and \citet{grasha2017}, where 
authors calculated the age difference in star cluster pairs, 
we can use colour index $U-B$, corrected for Galactic 
extinction and inclination effects, as a direct age indicator. The age of 
H\,{\sc ii}~bubbles (H$\alpha$ sources) is typically equal to a few Myr 
\citep{whitmore2011}. The age of star clusters without H$\alpha$ emission 
is tens and hundreds Myr (see Figs.~\ref{figure:ubv_sep} and 
\ref{figure:ubv_age}). Thus $\Delta t \equiv 
t({\rm SC})-t({\rm H}{\textsc {ii}}{\rm R})\approx t({\rm SC})$.

\begin{figure}
\vspace{0.7cm}
\resizebox{1.00\hsize}{!}{\includegraphics[angle=000]{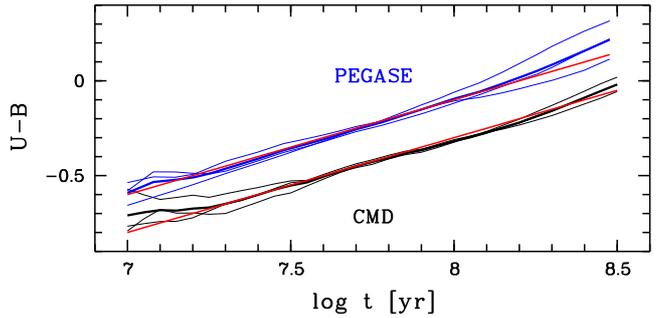}}
\caption{Variation of $U-B$ colour index of synthetic stellar systems vs. 
the age. Black thin curves are evolutionary tracks of CMD SSP stellar 
systems with $Z=$ 0.008, 0.012, and 0.019. Dark thin curves are evolutionary 
tracks of PEGASE SSP stellar systems with $Z=$ 0.008, 0.02, and 0.05. Thick 
black and dark curves are evolutionary tracks averaged over $Z$ for CMD and 
PEGASE models, respectively. Thick grey lines represent approximations 
$U-B\sim0.5\log t$ for both models.
}
\label{figure:ubt}
\end{figure}

As we noted above, star clusters without H$\alpha$ emission are free from 
dust \citep{whitmore2011}. Thus, we can estimate age of clusters by their 
colour indices, corrected for Galactic extinction and inclination effects, 
using evolutionary models. Variations of $U-B$ colour index for ageing
synthetic stellar systems are shown in Fig.~\ref{figure:ubt}.

\begin{table*}
\caption[]{\label{table:coeff}
Coefficients of equation~(\ref{equation:ts_gen}) for SC-H{\sc ii}R 
pairs and velocities of propagation of star formation.
}
\begin{center}
\begin{tabular}{cccccccccc} \hline \hline
Galaxy & \multicolumn{9}{c}{Range (pc)} \\
    & \multicolumn{3}{c}{40-200} & 
\multicolumn{3}{c}{40-500} & \multicolumn{3}{c}{100-500} \\
     & $\alpha$ & $\beta$ & $V_{\rm med}$ & $\alpha$ & $\beta$ & 
$V_{\rm med}$ & $\alpha$ & $\beta$ & $V_{\rm med}$ \\
& & & (km\,s$^{-1}$) & & & (km\,s$^{-1}$) & & & (km\,s$^{-1}$) \\
\hline
NGC  628 &  $1.18\pm0.13$ & $5.40\pm0.27$ & $3.0\pm1.9$ & $0.95\pm0.09$ 
& $5.83\pm0.20$ & $1.6\pm0.8$ & $0.59\pm0.19$ & $6.68\pm0.42$ & ... \\
NGC 3184 &  $1.03\pm0.27$ & $5.06\pm0.55$ & $8.3\pm1.1$ & $0.90\pm0.16$ 
& $5.20\pm0.35$ & $7.6\pm6.3$ & $0.91\pm0.36$ & $5.20\pm0.83$ & ... \\
NGC 3726 & $-0.40\pm0.32$ & $7.50\pm0.65$ & ... &         $1.22\pm0.17$ 
& $4.37\pm0.37$ & $29\pm25$   & $2.04\pm0.26$ & $2.38\pm0.62$ & ... \\
NGC 5585 &  $1.56\pm0.46$ & $4.09\pm0.97$ & ...   & $1.41\pm0.27$ 
& $4.38\pm0.62$ & ...         & $1.40\pm0.45$ & $4.39\pm1.05$ & ... \\
NGC 6946 &  $1.01\pm0.14$ & $6.00\pm0.27$ & $0.98\pm0.62$ & $0.76\pm0.10$ 
& $6.46\pm0.21$ & $0.54\pm0.27$ & $0.38\pm0.22$ & $7.35\pm0.48$ & ... \\
\hline
\end{tabular}
\end{center}
\end{table*}

Unfortunately, different modern evolutionary synthesis models do not provide
an unambiguous dependence between colour and age. As an example, we 
present two evolutionary models in Fig.~\ref{figure:ubt}. The first of 
them was described in Section~\ref{sect:select} (CMD), the second model 
was calculated using 
PEGASE.2\footnote{\url{http://www2.iap.fr/pegase/}} code \citep{fioc1997}. 
Both models are based on the Padova stellar library, both have the 
same parameters of stellar population (SSP, Salpeter IMF with the 
range of masses from $0.1 M_{\odot}$ to $100 M_{\odot}$). 
Fig.~\ref{figure:ubt} shows the similar behaviour of evolutionary tracks, 
however we observe the offset $\approx0.2$ mag between tracks of 
CMD and PEGASE models.

Given the uncertainties associated with the choice of a model, unknown 
metallicities of star clusters, and observational errors, 
we use a linear regression for the colour-age relation (thick grey lines 
in Fig.~\ref{figure:ubt}): $(U-B)_0^i\sim0.5\log t$. Evolutionary tracks 
with sub-solar and solar metallicities in both models satisfy this 
regression in the range of ages from $\approx15$ to $\approx160$ Myr. 
Based on more modern CMD model, we calculated ages of star 
clusters using the relation $\log t {\rm [yr]} = 2(U-B)_0^i+8.6$. 
The linear regression between age and colour allows us to minimize 
errors associated with photometric calibration for the $\log t - \log S$ 
slope estimation.

We used several key separation values. Two of them are characteristic 
scales of hierarchical star formation (100 pc is a typical size of 
OB-associations and 500 pc is a diameter of star complexes). The other 
two boundary values of $S$ were selected from the observations (40 pc 
is a minimal linear scale of our observational resolution and 200 pc 
is a maximal distance where we observe the correlation between the 
colour and the separation visually in Fig.~\ref{figure:col_dist}).

As a result, we chose three separation ranges for the further study of the 
$\log S - \log t$ dependence: an inner scale from 40 to 200 pc, an outer 
scale from 100 to 500 pc, and a general scale from 40 to 500 pc. 

\begin{figure}
\vspace{0.6cm}
\hspace{3mm}
\resizebox{0.85\hsize}{!}{\includegraphics[angle=000]{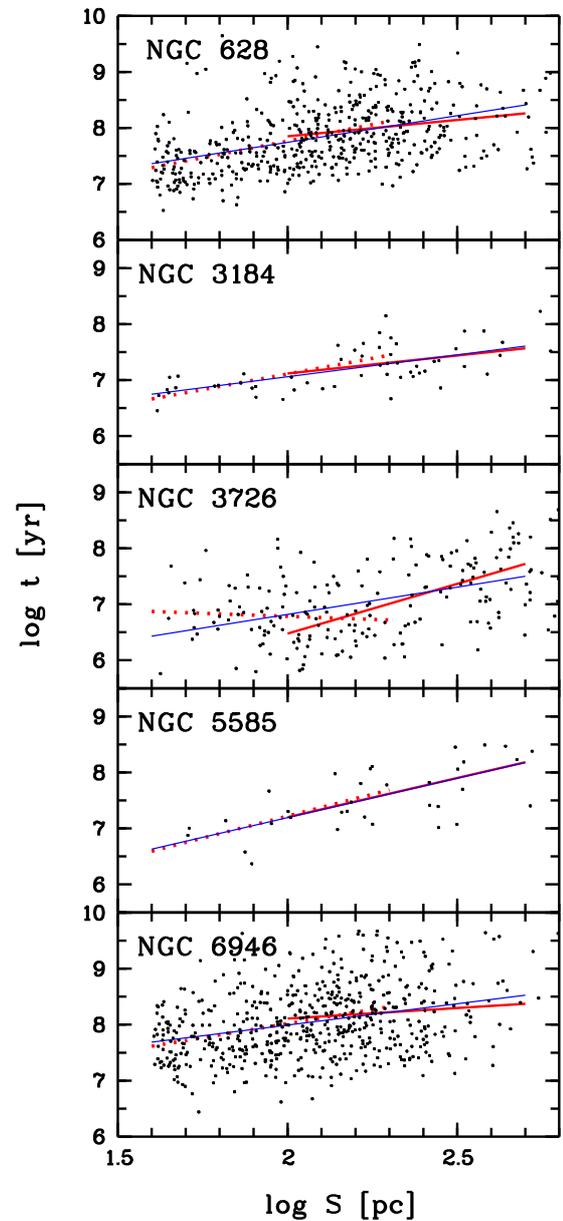}}
\caption{Logarithm of age of star clusters as a function of separation 
for SC-H{\sc ii}R pairs. The lines show linear regressions with 
coefficients from Table~\ref{table:coeff} for samples of 
SC-H{\sc ii}R pairs in the separation ranges 40-200 pc (dotted red lines), 
40-500 pc (blue lines), and 100-500 pc (solid red lines).
}
\label{figure:age_sep}
\end{figure}

Fig.~\ref{figure:age_sep} shows the dependence between $\log t$ and 
$\log S$ for star clusters in five studied galaxies in the used 
ranges of separation. Coefficients of the equation
\begin{equation}
\log t {\rm [yr]} = \alpha \log S {\rm [pc]}+\beta
\label{equation:ts_gen}
\end{equation}
for cluster samples in the galaxies are given in Table~\ref{table:coeff}.

Note that three of five studied galaxies (NGC 628, NGC 3184, and 
NGC 6946) have a regular symmetrical morphology. NGC 5585 has rather 
the Large Magellanic Cloud type. NGC 3726 has a distorted asymmetric 
disc. In addition, it is the most distant galaxy of the sample 
and, therefore, we were unable to identify a sufficient number of 
close SC-H{\sc ii}R pairs with $S<80$ pc in it (Fig.~\ref{figure:age_sep}).

The most reliable results are those obtained for NGC 628 and NGC 6946. 
On the one hand, these galaxies have a symmetrical regular structure, on 
the other hand, they provide us with a large statistical material for 
analysis -- more than 500 SC-H{\sc ii}R pairs.

The graphs in Fig.~\ref{figure:age_sep} and the coefficients given in 
Table~\ref{table:coeff} show closely resembling slopes for cluster samples 
in three galaxies (NGC 628, NGC 3184, and NGC 6946) with a regular 
structure. The calculated slopes lie in the range $1.0-1.2$ for separations 
from 40 to 200 pc. In the outer range of $S$ (100-500 pc), the slopes 
$\alpha\approx0.5$ for pair samples in NGC 628 and NGC 6946. This slope 
turned out to be larger in NGC 3184 ($\alpha=0.91$), however, it is 
calculated with poor accuracy due to insufficient statistics. 
The mean slope at 40 pc $<S<$ 500 pc is a little less than 1: 
$\alpha=0.85\pm0.10$ for SC-H{\sc ii}R pairs in all three galaxies 
(Table~\ref{table:coeff}). Coefficients $\beta$ differ from galaxy to 
galaxy. As we talked above, they are very sensitive to observational 
errors and model uncertainties.

The dependence in the peculiar galaxy NGC 5585 shows the approximately 
constant slope $1.4-1.6$ at all separation ranges. Unfortunately, 
the coefficient $\alpha$ is determined with a very large error.

SC-H{\sc ii}R pairs in the asymmetric galaxy NGC 3726 show a rather chaotic 
distribution on the diagram $\log t - \log S$. This scatter can be 
caused by various parameters of the medium in the asymmetric disc of 
the galaxy. Nevertheless, the mean slope at a general range of $S$ is 
equal to 1.22, that is close to the mean slope in NGC 5585.

Plots in Fig.~\ref{figure:age_sep} show that the typical age of star 
clusters in the range, that the power law is calculated over, 40-500 pc, 
is from 10 to 100-300 Myr. This result is in good agreement with 
previous studies of dependence between age difference and separation 
of cluster pairs, which gave correlations for star clusters with ages 
up to $\sim100$ Myr \citep{efremov1998,grasha2017}.

\subsection{Velocity of propagation of star formation}

Data obtained in Section~\ref{sect:age_sep} enable us to directly 
determine variations of star formation propagation velocity: 
$V = dS/dt$. The velocity is
\begin{equation}
V{\rm [km\,s^{-1}]} = 
\gamma^{-1/\alpha}\alpha^{-1}t{\rm [Myr]}^{1/\alpha-1}
\label{equation:v1t}
\end{equation}
or 
\begin{equation}
V{\rm [km\,s^{-1}]} = (\alpha\gamma)^{-1}S{\rm [pc]}^{1-\alpha},
\label{equation:v2s}
\end{equation}
where $\gamma=10^{\beta-6}$, and $\alpha$, $\beta$ are coefficients from 
equation~(\ref{equation:ts_gen}).

We calculated median velocities, $V_{\rm med}$, for 
SC-H{\sc ii}R pairs using equations~(\ref{equation:v1t}), 
(\ref{equation:v2s}). These values are presented in 
Table~\ref{table:coeff}. We also estimated velocity errors, $\Delta V$, 
based on the coefficients $\alpha$, $\beta$, and their uncertainties 
from Table~\ref{table:coeff}. Significant part of SC-H{\sc ii}R pair 
samples has extremely large velocity estimation errors, 
$\Delta V > V_{\rm med}$. These estimates are not shown in the table.

The data show the approximately constant or slightly falling velocity 
of star formation propagation in three regular structure galaxies 
(NGC 628, NGC 3184, and NGC 6946) with typical velocities 
$V\sim1-10$ km\,s$^{-1}$ at separation scales from 40 to 200 pc. 
The velocity of star formation propagation falls more steeply with 
time and distance in peculiar galaxies NGC 3726 and NGC 5585 
at all studied ranges of $S$. The exception is the inner range 40-200 pc 
in NGC 3726. As we noted above, this result is unreliable due to 
a small number of close SC-H{\sc ii}R pairs in the galaxy. Nevertheless, 
typical velocities $V$ in NGC 3726 and NGC 5585 are also close to a few 
km\,s$^{-1}$ as in three galaxies with regular structure.

Variations of $V$ and $\alpha$ for different ranges of $S$ and different 
morphology of galaxies will be discussed in the next section.

\section{Discussion}

\citet{whitmore2011} studied young star clusters in the grand-design galaxy 
M83 using the morphology of the surrounding H$\alpha$ emission. The 
relation between the age of a star cluster and radius of H$\alpha$ bubble, 
found in \citet{whitmore2011}, showed a strong age-size correlation
for star clusters younger than $\log (t/{\rm yr})=6.7-6.8$, 
with ionized gas bubble growing from a few to $\approx20$ pc in size. 
There is no age-size correlation for star clusters older than
$\log (t/{\rm yr})=6.8$ and bubble's radius $>20$ pc. At this distance, 
the gas bubble detaches from the star cluster and begins to blur. 
Its expansion is weakly dependent on the central energy source 
(explosions of supernovae). The average age of star clusters in M83 at 
this distance (20 pc) is $\log (t/{\rm yr})=6.76$. Our data obtained 
from equation~(\ref{equation:ts_gen}), and coefficients in 
Table~\ref{table:coeff} for clusters in NGC 628 (an example of a 
galaxy which is similar to M83 by luminosity and morphology) in the range 
20-400 pc, give a close age $\log (t/{\rm yr})=6.9\pm0.3$ at a distance 
of 20 pc.

\begin{figure}
\vspace{0.7cm}
\resizebox{0.95\hsize}{!}{\includegraphics[angle=000]{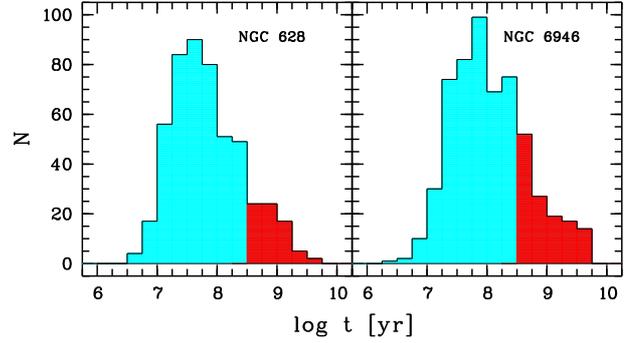}}
\caption{Distribution of star clusters in NGC 628 and NGC 6946 by age. 
Clusters older than 300~Myr are marked by red. See the text for details.
}
\label{figure:classic}
\end{figure}

\begin{figure}
\vspace{0.7cm}
\hspace{2mm}
\resizebox{0.90\hsize}{!}{\includegraphics[angle=000]{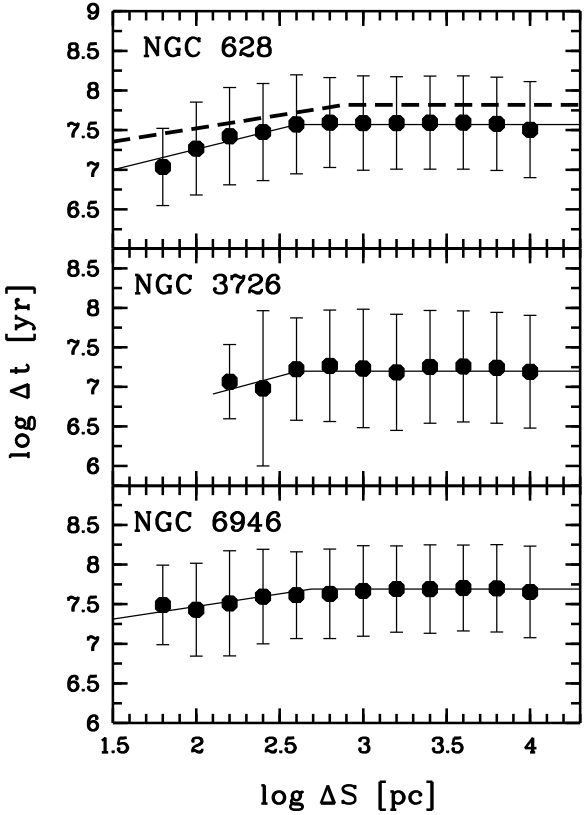}}
\caption{Age difference vs. separation for cluster pairs in NGC 628, 
NGC 3726, and NGC 6946. The black circles and error bars show the average 
age and $1\sigma$ spread for each cluster pair in each 0.2~dex separation
bin. Solid thin lines show linear regression at smaller separations and
flat regression at larger separations to the 
$\log \Delta t - \log \Delta S$ relation for the galaxies from our
data. Thick dashed line is the regression to the age-separation
relation for NGC 628 according to \citet{grasha2017}. See the text
for details.
}
\label{figure:classic2}
\end{figure}

\begin{figure}
\vspace{0.7cm}
\hspace{2mm}
\resizebox{0.90\hsize}{!}{\includegraphics[angle=000]{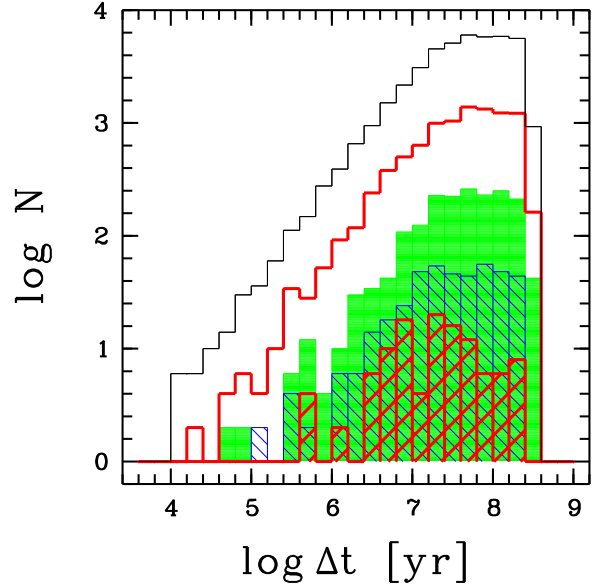}}
\caption{Age difference distributions of cluster pairs in 
NGC 628 with $\log \Delta S = 3.6\pm0.1$ (black histogram), 
$3.2\pm0.1$ (red), $2.8\pm0.1$ (green filled), $2.4\pm0.1$ (blue shaded), 
and $2.0\pm0.1$ (red shaded). See the text for details. 
}
\label{figure:hist628}
\end{figure}

\begin{figure}
\vspace{0.8cm}
\hspace{3mm}
\resizebox{0.90\hsize}{!}{\includegraphics[angle=000]{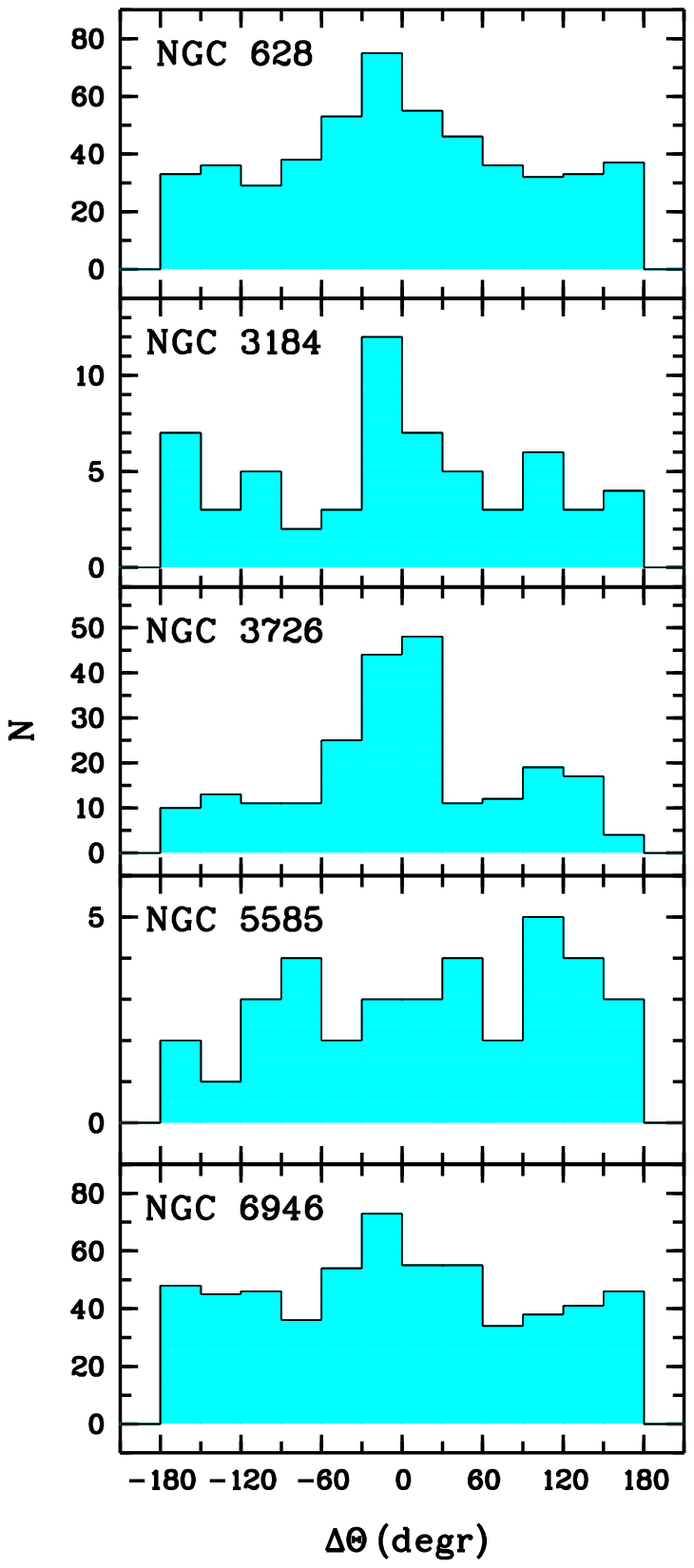}}
\caption{Distribution of SC-H{\sc ii}R pairs by $\Delta\Theta$. See 
the text for details.
}
\label{figure:angle}
\end{figure}

The data for other four galaxies (NGC 3184, NGC 3726, 
NGC 5585, and NGC 6946) at the ranges of $S$ 40-200 pc and 40-500 pc 
give $\log(t/{\rm yr})=6.2-7.4$ at the distance 20 pc. Unfortunately, 
the errors of $\log(t/{\rm yr})$ exceed 0.4~dex for all galaxies except 
NGC 6946.

We applied the standard technique for studying an extension of 
hierarchical star formation \citep{efremov1998,grasha2017} to our data. 
Separation $\Delta S$ and age difference $\Delta t$ between all identified 
cluster pairs based on their coordinates and $U-B$ colour photometric 
ages were calculated. Our star clusters are randomly distributed in the
galactic plane, thus the number of pairs with separation 
$\Delta S<500$ pc is too little to use statistical methods for galaxies 
NGC 3184 and NGC 5585. Therefore we analyse dependence 
$\log \Delta t - \log \Delta S$ for cluster pairs only in three 
galaxies: NGC 628, NGC 3726, and NGC 6946.

Fig.~\ref{figure:classic} shows the age distribution of star 
clusters in NGC 628 and NGC 6946. As seen from the histograms, star 
clusters are distributed by Gauss law with a little surplus of old 
clusters ($t>1$ Gyr). We excluded star clusters older than 300 Myr (they 
are marked by red colour in Fig.~\ref{figure:classic}) from the study of 
the $\log \Delta t - \log \Delta S$ dependence of cluster pairs, 
following \citet{grasha2017}. We do not show here the histogram for star 
clusters in NGC 3726, because of absence of clusters older than 300 Myr in 
the galaxy (see Fig.~\ref{figure:age_sep}).

Obtained dependences for cluster pairs in NGC 628 and NGC 6946 resemble 
dependences between age difference and separation of cluster pairs 
found earlier in \citet{efremov1998} and \citet{grasha2017}. We observe 
a growth of $\Delta t$ with increase in $\Delta S$ with slopes $0.52\pm0.12$ 
in NGC 628, $0.57\pm0.47$ in NGC 3726, and $0.30\pm0.11$ in NGC 6946 
till the separations $\approx400-500$ pc in the galaxies. These 
distributions reach a plateau at $\Delta S \sim400-500$ pc in the galaxies 
(Fig.~\ref{figure:classic2}).

The obtained slopes are in close agreement with those previously found by
\citet{grasha2017}. We can compare $ \Delta t - \Delta S$ relation for 
cluster pairs in NGC 628 as this galaxy is common for both our sample 
and the sample studied by \citet{grasha2017}. 
The result of \citet{grasha2017}: 
$\log \Delta t \sim (0.33\pm0.07)\log \Delta S$ for $\Delta S<600-950$ pc 
is consistent within errors with our fitting (see the thick dashed 
line, black circles, error bars, and the solid thin line in the top 
panel of Fig.~\ref{figure:classic2}). This is an additional argument in 
support of our cluster age estimates via $U-B$ colour.

The range of spatial scales for the galaxies in 
Fig.~\ref{figure:classic2} is larger compared to Fig.~\ref{figure:age_sep}. 
This is due to the fact that graphs in Fig.~\ref{figure:classic2} 
show separations between all star clusters located randomly in the disc, 
whereas graphs in Fig.~\ref{figure:age_sep} show separations of star clusters 
from H\,{\sc ii} regions nearest to them. As a result, the spatial scale of 
cluster pairs in Fig.~\ref{figure:classic2} corresponds to the spatial 
scale of a galactic disc, whereas the spatial scale of SC-H{\sc ii}R pairs 
corresponds to maximal distances between star clusters and H\,{\sc ii} 
regions nearest to them. This spatial scale corresponds to a typical size 
of a star complex.

Of course, our estimates of the $\log \Delta t - \log \Delta S$ 
dependency are only qualitative. They are intended to show an agreement 
between our estimates and the results of previous researchers 
\citep{efremov1998,grasha2017}, if we use the classical technique to 
study the hierarchical star formation in turbulent media based on the study of
the $\log \Delta t - \log \Delta S$ dependence for cluster pairs. 
\citet{efremov1998} and \citet{grasha2017} used cluster ages determined 
directly. Obviously, our 'photometric' ages, calculated from the $U-B$ 
colour, are less reliable.

We did not identify close cluster pairs ($S<100$ pc) in NGC 3726, the 
most distant galaxy in the sample. Thus, correlation parameters 
for it were obtained with large errors and are shown in the figure only 
for a qualitative comparison.

Our results show that the decrease of the mean age difference with 
the decrease of separation $\Delta S$ for close cluster pairs with 
$\Delta S<250$ pc is a real physical effect. Age difference distributions 
for cluster pairs within narrow $\Delta t$ ranges (0.2~dex), which are 
presented in Fig.~\ref{figure:hist628}, show similar shapes for 
pairs with $\log \Delta S\ge2.8$. However, the distribution of close 
cluster pairs ($\log \Delta S=2.0$) differs significantly from them. 
A relative deficiency of cluster pairs with a large age difference is 
observed clearly for pairs with $\log \Delta S=2.0\pm0.1$ (the red 
shaded histogram in the figure).

On large spatial scales ($\log \Delta S>2.4$), distributions of 
cluster pairs reach a maximum $N_{\rm max}$ at 
$\log \Delta t \approx 7.5$ and have a flat profile for the pairs 
with $\log \Delta t = 7.5-8.4$. Note that 
$\log N_{\rm max} \sim \log \Delta S$ (Fig.~\ref{figure:hist628}). 
These facts indicate a significant proportion of cluster pairs with 
$\Delta t \sim 100$ Myr, randomly and evenly distributed on spatial scales 
of a few hundred pc and more. On smaller scales ($\le100$ pc), the 
existence of cluster pairs with large age difference is difficult. 
Starburst at $\sim100$ pc scale has a duration of $\sim10$ Myr 
\citep{efremov1998}, it destroys H$_2$ cloud remnants. Star formation 
here becomes impossible for at least 100 Myr (ISM relaxation time).

Stationary density wave theory predicts the existence of an age gradient 
across the spiral arms at least in grand-design galaxies 
\citep[see survey in][]{dobbs2014}. Observational studies show this 
gradient in some galaxies \citep[see][and references therein]{shabani2018}. 
We checked if there is an age gradient across arms based on our data.
For this purpose, we constructed angle $\Delta\Theta$ distributions, where 
$\Delta\Theta$ is the angle between the position angle of a star cluster 
and the position angle of the H\,{\sc ii} region relative to this cluster 
(Fig.~\ref{figure:angle}). If SC-H{\sc ii}R pair is located strictly 
across a spiral arm, and H\,{\sc ii} region is located in the outer part 
of the arm, $\Delta\Theta$ is equal to the pitch angle of the spiral arm, 
typically $10-20\degr$.

Distributions of SC-H{\sc ii}R pairs by $\Delta\Theta$ in studied 
galaxies show a strong peak near $\Delta\Theta=0\degr$ in NGC 3184 and 
NGC 3726. Both galaxies have grand-design morphology. Weak peaks are 
observed in NGC 628 and NGC 6946. NGC 5585 does not have a preferred 
direction of SC-H{\sc ii}R pairs across spiral arms. This is not 
surprising for peculiar NGC 5585. Our results support the result of 
\citet{shabani2018}, who did not find a noticeable age gradient across 
spiral arms in NGC 628. We identified star clusters from catalogue of 
\citet{shabani2018} and compared the obtained ages for common objects. 
Estimates showed a good agreement between our ages, $t$, and ages found 
in \citet{shabani2018}, $t_{\rm ref}$: 
$\log t = \log t_{\rm ref}+0.09\pm0.38$.

Notice that presence of large star complexes must smooth the effect 
of age gradient across spiral arms. Star formation begins in central, 
densest and coldest part of GMC, and it propagates toward outer parts. 
As a result, we must observe older star formation regions in the 
central part of a star complex. NGC 628 is just an example of a galaxy 
with numerous star complexes \citep{gusev2013}.

The direction of star formation propagation within star complexes 
are indirectly confirmed by the fact that the maximal distance where we 
observe the correlation between age of a star cluster and 
SC-H{\sc ii}R separation is about 500 pc in all studied galaxies 
(Fig.~\ref{figure:col_dist}). This is the typical radius of
star complexes \citep{efremov1998}. 

The calculated coefficients $\beta$ in equation~(\ref{equation:ts_gen}) 
for the galaxies differ by order of magnitude. As we discussed in 
Section~\ref{sect:age_sep}, they are very sensitive to observational 
errors and uncertainties in models. If the colour index $U-B$ is shifted 
by 0.1~mag, the constant in equation~(\ref{equation:ts_gen}) will change 
by 0.2~dex. Additionally, there are physical reasons for the difference in 
the coefficient $\beta$, which is responsible for the value of star 
formation propagation velocity $V$, caused by difference in integral 
parameters of galaxies. As is known, in smaller galaxies the star 
formation length- and time-scales are generally shorter than in large 
galaxies \citep{elmegreen1996b}. We can only assume that velocities 
of star formation propagation are equal to a few km\,s$^{-1}$ in order 
of magnitude. This does not contradict results obtained by previous 
researchers \citep[see][for instance]{grasha2017}.

Slopes $\alpha$ of the age-separation dependency, obtained in 
Section~\ref{sect:age_sep}, vary with the range of separations and the 
galactic morphology. These variations reflect different contributions 
of different physical processes occurring in the interstellar medium 
(ISM). On spatial scales of 100-500 pc, slope $\alpha$ approaches 
0.5 in the galaxies with a regular structure. This result confirms 
conclusions of previous observational and theoretic studies 
\citep{efremov1998,grasha2017,nomura2001} on the dominant role 
of turbulence in propagation of the star formation process on spatial 
scales up to 500 pc. On a smaller scale, $S\le200$ pc, the slope $\alpha$ 
in these galaxies is equal to $\approx1$. It indicates a significant 
contribution of other physical processes, such as stellar winds and 
supernova explosions, to propagation of the star formation process. 
The expansion of the H\,{\sc ii} region (Str\"{o}mgren sphere) 
satisfies the expression $t \sim R^{1.75}$, where $t$ is time and $R$ is 
radius. Obtained expansion law $t \sim S^{1}$ may involve, along 
with turbulence, complications from stellar winds and probably supernovae, 
which the simple Str\"{o}mgren formula does not \citep{elmegreen2019}.

This complex mechanism seems to work in galaxies with an asymmetric 
structure at large spatial scales, up to 500 pc. The slope $\alpha>1$ 
and it is systematically larger than in galaxies with a regular structure 
(see Table~\ref{table:coeff}). This fact may indicate a smaller contribution 
of turbulence in the process of star formation propagation in the 
perturbed interstellar medium of these galaxies.

Observations of resolved stellar populations in nearest galaxies, such 
as LMC \citep{sun2018}, allow as to explore in detail a hierarchical star 
formation and an influence of turbulence in galactic discs on a wide range of 
spatial scales. Parameters and correlations of young stellar structures in 
nearby galaxies (fractal dimension, size, density, and mass functions, 
selected scales of stellar structures) obtained by different authors are 
in a good agreement with each other. These parameters demonstrate properties 
which are similar to those of the ISM, which is regulated by supersonic 
turbulence. Some features are observed only in starburst dwarf galaxies 
\citep{elmegreen2014}.

The method for studying the SC-H{\sc ii}R pairs, presented in this paper, 
extends our ability to explore the process of hierarchical star formation in 
turbulent ISM to distant galaxies. The most interesting outcome, in our 
opinion, is the study of star formation propagation on scales of star 
associations ($\le100$ pc), where different physical processes, along with 
turbulence, seem to play a significant role.

\section{Conclusions}

We developed a method for investigating the process of star formation 
propagation in the turbulent interstellar medium based on the study of 
the separation of young star clusters from the nearest H\,{\sc ii} regions.
We found a relation between the separation $S$ and the age $t$ 
($U-B$ colour index) of star clusters on spatial scales from 40 to 
500 pc and on time scales from 10 to 300 Myr.

The average age of star clusters in the galaxies with symmetric 
morphology satisfies the relation $t\sim S^{1.0-1.2}$ in the 
separation range of 40-200 pc and $t\sim S^{0.4-0.9}$ in the range of 
100-500 pc. The galaxies with the asymmetric structure show steeper 
dependence between the age and the separation 
with a power $>1.2$ for separations from 40 to 500 pc.

Our results confirm the conclusions of previous studies on the dominant 
role of turbulence in propagation of the hierarchical star formation 
process on spatial scales from 200 to 500 pc in galaxies with 
symmetric structure. On spatial scales of $\le100$ pc in the galaxies 
with regular morphology, as in peculiar galaxies, other physical 
processes may contribute to the star formation propagation process 
along with turbulence.

The velocity of star formation propagation is nearly constant or 
falls weakly on the spatial scales of $<200$ pc and it has typical 
values of a few km\,s$^{-1}$.

\section*{Acknowledgments}
We are extremely grateful to the anonymous referee for his/her constructive 
comments. The authors would like to thank Yu.~N.~Efremov, A.~V.~Zasov, 
O.~K.~Silchenko, O.~V.~Egorov, I.~Yu.~Katkov, S.~B.~Popov (SAI MSU), 
and B.~G.~Elmegreen (IBM Research Division, T.~J.~Watson Research 
Center) for helpful discussions. AG acknowledges the support from the 
Program of development of Lomonosov Moscow State University (Leading 
Scientific School 'Physics of stars, relativistic objects and galaxies').


\begin{thebibliography}{}

\bibitem [Ali et al.(2017)]{ali2017}
          Ali K., Obreschkow D., Fisher D.B., Glazebrook K., Damjanov I., 
          Abraham R.G., Bassett R. 2017, ApJ, 845, id. 37

\bibitem [Dale et al.(2009)]{dale2009}
          Dale D.~A. et al. 2009, ApJ, 703, 517

\bibitem [de la Fuente Marcos \& de la Fuente Marcos(2009)]{marcos2009}
          de la Fuente Marcos R., de la Fuente Marcos C. 2009, 
          ApJ, 700, 436

\bibitem [Dobbs \& Baba(2014)]{dobbs2014}
          Dobbs C., Baba J. 2014, Publ. Astr. Soc. Australia, 
          31, id. e035

\bibitem [Efremov \& Elmegreen(1998)]{efremov1998}
          Efremov Y.~N., Elmegreen B. 1998, MNRAS, 299, 588

\bibitem [Elmegreen(2002)]{elmegreen2002}
          Elmegreen B.~G. 2002, ApJ, 564, 773

\bibitem [Elmegreen(2006)]{elmegreen2006}
          Elmegreen B.~G. 2006, in Del Toro Iniesta J.~C., 
          Alfaro E.~J., Gorgas J.~G., Salvador-Sole E., 
          Butcher H., eds, The Many Scales in the Universe:
          JENAM 2004 Astrophysics Reviews. Springer, Dordrecht, 
          p. 99

\bibitem [Elmegreen(2010)]{elmegreen2010}
          Elmegreen B.~G. 2010, in R. de Grijs \& J. L\'{e}pine, eds, 
          Proc. IAU Symp. 266, Star Clusters: Basic Galactic Building 
          Blocks Throughout Time and Space. Cambridge Univ. Press, 
          Cambridge, p. 3

\bibitem [Elmegreen(2011)]{elmegreen2011}
          Elmegreen B.~G. 2011, in Charbonnel C., Montmerle T., 
          eds, in EAS Publications Series 51, Ecole Evry Schatzman 2010: 
          Star Formation in the Local Universe. Cambridge Univ. Press, 
          Cambridge, p. 31

\bibitem [Elmegreen(2019)]{elmegreen2019}
          Elmegreen B.~G. 2019, private commun.

\bibitem [Elmegreen \& Efremov(1996)]{elmegreen1996}
          Elmegreen B.~G., Efremov Y.~N. 1996, ApJ, 466, 802

\bibitem [Elmegreen \& Lada(1977)]{elmegreen1977}
          Elmegreen B.~G., Lada C.~J. 1977, ApJ, 214, 725

\bibitem [Elmegreen et al.(1996)]{elmegreen1996b}
          Elmegreen B.~G, Elmegreen D.~M., Salzer J., Mann H. 1996, 
          ApJ, 467, 579

\bibitem [Elmegreen et al.(2000)]{elmegreen2000}
          Elmegreen B.~G., Efremov Y., Pudritz R.~E., Zinnecker H. 
          2000, in Mannings V., Boss A.~P., Russell S.~S., eds, 
          Protostars and Planets IV. Univ. of Arizona Press, Tucson, 
          p. 179

\bibitem [Elmegreen et al.(2014)]{elmegreen2014}
          Elmegreen D.~M. et al. 2014, ApJL, 787, id. L15

\bibitem [Fisher et al.(2017)]{fisher2017}
          Fisher D.B. et al. 2017, ApJL, 839, id. L5

\bibitem [Fioc \& Rocca-Volmerange(1997)]{fioc1997}
          Fioc M., Rocca-Volmerange B. 1997, A\&A, 326, 950

\bibitem [Grasha et al.(2015)]{grasha2015}
          Grasha K. et al. 2015, ApJ, 815, id. 93

\bibitem [Grasha et al.(2017)]{grasha2017}
          Grasha K. et al. 2017, ApJ, 842, id. 25

\bibitem [Grasha et al.(2019)]{grasha2019}
          Grasha K. et al. 2019, MNRAS, 483, 4707

\bibitem [Gusev \& Efremov(2013)]{gusev2013}
          Gusev A.~S., Efremov Yu.~N. 2013, MNRAS, 434, 313

\bibitem [Gusev et al.(2015)]{gusev2015}
          Gusev A.~S., Guslyakova S.~A., Novikova A.~P., Khramtsova M.~S.,
          Bruevich V.~V., Ezhkova O.~V. 2015, Astron. Rep., 59, 899

\bibitem [Gusev et al.(2016)]{gusev2016}
          Gusev A.~S. et al. 2016, MNRAS, 457, 3334

\bibitem [Gusev et al.(2018)]{gusev2018}
          Gusev A.~S., Shimanovskaya E.~V., Shatsky N.~I., Sakhibov F., 
          Piskunov A.~E., Kharchenko N.~V. 2018, Open Astronomy, 27, 98

\bibitem [James et al.(2004)]{james2004}
          James P.~A. et al. 2004, A\&A, 414, 23

\bibitem [Kennicutt et al.(2008)]{kennicutt2008}
          Kennicutt R.~C.~Jr, Lee J.~C., Funes J.~G.~J.~S., Sakai S.,
          Akiyama S. 2008, ApJS, 178, 247

\bibitem [Knapen et al.(2004)]{knapen2004}
          Knapen J.~H., Stedman S., Bramich D.~M., Folkes S.~L.,
          Bradley T.~R. 2004, A\&A, 426, 1135

\bibitem [Lada \& Lada(2003)]{lada2003}
          Lada C.~J., Lada, E.~A. 2003, ARA\&A, 41, 57

\bibitem [Larsen \& Richtler(1999)]{larsen1999b}
          Larsen S.~S., Richtler T. 1999, A\&A, 345, 59

\bibitem [Larson(1981)]{larson1981}
          Larson R.~B. 1981, MNRAS, 194, 809

\bibitem [Marigo et al.(2008)]{marigo2008}
          Marigo P., Girardi L., Bressan A., Groenewegen
          M.~A.~T., Silva L., Granato G.~L. 2008, A\&A, 482, 883

\bibitem [Nomura \& Kamaya(2001)]{nomura2001}
          Nomura H., Kamaya H. 2001, AJ, 121, 1024

\bibitem [Paturel at al.(2003)]{paturel2003}
          Paturel G., Petit C., Prugniel Ph., Theureau G., Rousseau J.,
          Brouty M., Dubois P., Cambresy L. 2003, A\&A, 412, 45

\bibitem [Piskunov at al.(2011)]{piskunov2011}
          Piskunov A.~E., Kharchenko N.~V., Schilbach E., R\"{o}ser S., 
          Scholz R.-D., Zinnecker H. 2011, A\&A, 525, 122

\bibitem [Scalo(1985)]{scalo1985}
          Scalo J.~M. 1985, in D.~C. Black \& M.~S. Mathews, eds, 
          Protostars and Planets II. Univ. Arizona Press, Tucson, p. 201

\bibitem [Shabani et al.(2018)]{shabani2018}
          Shabani F. et al. 2018, MNRAS, 478, 3590

\bibitem [Sun et al.(2018)]{sun2018}
          Sun N.-C. et al. 2018, ApJ, 858, id. 31

\bibitem [Whitmore et al.(2010)]{whitmore2010}
          Whitmore B.~C. et al. 2010, AJ, 140, 75

\bibitem [Whitmore et al.(2011)]{whitmore2011}
          Whitmore B.~C. et al. 2011, ApJ, 729, 78

\end{thebibliography}
\end{document}